\documentclass[showpacs,preprintnumbers,
amsmath,amssymb,aps,prd,nofootinbib,superscriptaddress,cite
eqsecnum,11pt]{revtex4}
\usepackage{graphicx,amsmath, amssymb, graphics,ascmac,fancybox}
\usepackage{color,booktabs,fancyhdr,bm,comment}

\makeatletter
\renewcommand{\p@subsection}{}
\makeatother

\makeatletter
\newcommand{\xequal}[2][]{\ext@arrow 0055{\equalfill@}{#1}{#2}}
\def\equalfill@{\arrowfill@\Relbar\Relbar\Relbar}
\makeatother

\newcommand{\chushi}[1]{ }

\newcommand{\bard}[1]{ \overline{ #1 } }

\newcommand{\angleLR}[1]{ \left \langle #1 \right \rangle }
\newcommand{\angleN}[1]{ \langle #1 \rangle }

\newcommand{\roundLR}[1]{ \left( #1 \right) }
\newcommand{\roundN}[1]{ ( #1 ) }

\newcommand{\roundB}[1]{ \biggl( #1 \biggr) }

\newcommand{\squareLR}[1]{ \left[ #1 \right] }

\newcommand{\squareB}[1]{ \biggl[ #1 \biggr] }
\newcommand{\squareBB}[1]{ \Biggl[ #1 \Biggr] }

\newcommand{\absLR}[1]{ \left| #1 \right| }

\newcommand{\eq}[1]{\begin{equation}\begin{split} #1 \end{split}\end{equation}}

\newcommand {\mathsym}[1]{{}}
\newcommand {\unicode}[1]{{}}

\makeatletter

\@addtoreset{equation}{section}
\makeatother

\begin{document}

\title{Effective temperature of nonequilibrium dense matter in holography}

\author{Hironori Hoshino}
\affiliation{
Department of Physics, Nagoya University,
Nagoya 464-8602, Japan
}
\author{Shin Nakamura}
\affiliation{
Department of Physics, Chuo University,
Tokyo 112-8551, Japan
}
\affiliation{
Institute for Solid State Physics, University of Tokyo,
Kashiwa 277-8581, Japan
}


\begin{abstract}
We study properties of effective temperature of nonequilibrium steady states by using the anti-de Sitter spacetime/conformal field theory (AdS/CFT) correspondence. We consider nonequilibrium systems with a constant flow of current along an electric field, in which the current is carried by both the doped charges and those pair created by the electric field. We find that the effective temperature agrees with that of the Langevin systems if we take the limit where the pair creation is negligible. The effect of pair creation raises the effective temperature whereas the current by the doped charges contributes to lower the effective temperature in a wide range of the holographic models. 
\end{abstract}
\maketitle
\section{introduction}
\label{introduction}
Nonequilibrium physics is one of the frontiers of modern physics, and construction of nonequilibrium statistical mechanics is still a great challenge. 
The difficulty comes from the fact that we cannot \textit{a priori} rely on the guiding principle, such as the principle of detailed balance, held in equilibrium systems. However, we wish to find a fundamental law that governs a wide range of nonequilibrium systems, or more precisely, we wish to know if such a fundamental law exists or not. A good place to study for this purpose may be nonequilibrium steady states (NESS). In NESS, the system is driven by a constant external force, and the system is out of equilibrium with dissipation, but the macroscopic variables do not evolve in time. 

Recently, the anti-de Sitter spacetime/conformal field theory (AdS/CFT) correspondence, which is a computational
method developed in superstring theory, has been
applied to studies of out-of-equilibrium physics (see, for example, \cite{Hubeny:2010ry}). In the framework of AdS/CFT, that is also
called holography, a physical problem can be reformulated
into that in gravity theory, and one finds that the original
problem can be much more easily analyzed in the gravity
dual.
For example, transport coefficients in NESS have been computed beyond the linear response regime~\cite{Herzog:2006gh,Gubser:2006bz,Karch:2007pd}. The typical systems of NESS are the systems of a test particle dragged at a constant velocity in a medium~\cite{Herzog:2006gh,Gubser:2006bz} (which we call Langevin systems in this paper) and systems of charged particles with constant flow of current along the external electric field acting on the charge~\cite{Karch:2007pd} (which we call conductor systems).

It has also been found that the notion of the effective temperature of NESS in these systems naturally appears in the gravity dual picture in terms of the Hawking temperature of an analogue black hole~\cite{Gubser:2006nz,CasalderreySolana:2007qw,Gursoy:2010aa,Kim:2011qh,Sonner:2012if,Nakamura:2013yqa}. 
The effective temperature agrees with the ratio between the fluctuation and the dissipation at NESS~\cite{Gursoy:2010aa,Nakamura:2013yqa} and it characterizes the correlation functions of fluctuations in NESS\footnote{For the definition of the effective temperature in the literature
on nonequilibrium statistical physics, see, for example, \cite{effectiveT} for a review.}. Therefore, the effective temperature is quite important in the research into nonequilibrium statistical physics. 

In this paper, we further study the nature of the effective temperature in holographic models. One of the problems we shall study in this paper is the relationship between the effective temperature of the conductor systems and that in the Langevin systems. Since the conductor systems consist of many charged particles, their effective temperature may be related to that in the Langevin systems where a single (but the same) charged particle is dragged. In general, the effective temperatures of these two systems are different from each other. However, as we shall see, if we take the large-mass limit or the large-density limit of the charge carriers in the conductor systems, the effective temperature agrees with that in the Langevin systems.
In order to reach the aforementioned results, we introduce the mass of the charged particles and the charge density to the analysis of \cite{Nakamura:2013yqa}, where the zero-mass and zero-density limits have been taken. In \cite{Nakamura:2013yqa}, it has been found that the effective
temperature of NESS can be either higher or lower than the
temperature of the heat bath, depending on the models and
the parameters of the systems. At finite densities, we shall find that the effective temperature can be lower than the temperature of the heat bath even for the models that had the higher effective temperatures at zero density in \cite{Nakamura:2013yqa}. Our results imply that the effect of pair creation of charge
carriers by the external force is responsible for raising the
effective temperature, whereas the effect of dragging the
already-existing doped charge carriers is responsible for
lowering the effective temperature, at least for our systems.

The organization of the paper is as follows. In Section \ref{setup}, we overview previous works on computation of nonlinear conductivity in holographic models. The setup of our
model is also explained. One representative of the holographic
models of conductors is the so-called D3-D7 model~\cite{Karch:2007pd}. Therefore, we mainly focus on the D3-D7 model in this paper. However, the computations are straightforwardly generalized into other models.  In Section \ref{fluctuations}, the derivation of
effective temperature is presented. Our main results shall be given in Section \ref{results}, and we conclude in Section \ref{discussions}.

\section{Setup}
\label{setup}

In order to prepare NESS, we need a subsystem coupled to a heat bath. We apply an external force which drives the subsystem into nonequilibrium, and the heat bath absorbs the dissipation. When the work given by the external force and the dissipation into the heat bath are in balance, the subsystem can be realized as NESS. For the conductor systems in this study, the external force is the external electric field $E$, the subsystem is a many-body system of charge carriers, and the heat bath is a system of particles that are neutral in $E$ and are interacting with the subsystem.

One typical realization of the foregoing conductor system in holography is the D3-D7 system with electric field~\cite{Karch:2007pd}. Since the analysis can be straightforwardly generalized into the cases of other models, we mainly focus on the D3-D7 system in this paper. Let us briefly review the model of~\cite{Karch:2007pd} to explain our setup and notations.

The field theory realized on the D3-D7 system is a supersymmetric QCD that consists of a $(3+1)$-dimensional $SU(N_{c})$ ${\cal N}=4$ supersymmetric Yang-Mills (SYM) theory for adjoint representations (which we call ``gluons'') and a ${\cal N}=2$ hypermultiplet as a sector of fundamental representations (which we call ``quarks'' or ``antiquarks''). We apply an external force electrically acting on the quark charge (which we call ``electric field''), and then the current of the quark charge (which we call ``current'') appears. The gluon sector plays the role of a heat bath since it absorbs the dissipation produced in the quark sector. The picture of the heat bath is established since we take the large-$N_{c}$ limit where  the degrees of freedom of the gluon sector are infinitely large compared to that of the quark sector. We also take the large 't Hooft coupling limit so that the typical interaction scale of the microscopic process is short enough compared to that of the macroscopic physics.

In the gravity dual picture, the heat bath of the gluon sector is mapped to the geometry of a direct product of an AdS-Schwarzschild black hole (AdS-BH) and an $S^5$, whose metric is given by
\eq{
	ds^2
&=
	\hat{g}_{\mu \nu}
	dx^\mu
	dx^\nu
=
	-
	\frac{1}{z^2}
	\frac{(1 - z^4/z^4_H)^2}{1 + z^4/z^4_H}
	dt^2
	+
	\frac{1}{z^2}
	\roundLR{1 + z^4/z^4_H}
	d\vec{x}^2
	+
	\frac{dz^2}{z^2}
	+
	d \Omega^2_5
,
\label{eq:metricofmetallic}
}
where we have set the AdS radius to be 1.
$z$ is the radial coordinate on which the horizon is located at $z=z_{H}$, and the boundary is at $z=0$. The Hawking temperature is given by $T=\frac{\sqrt{2}}{\pi z_H}$. The boundary extends along $(t, \Vec{x})$ directions, and $d\Omega_{5}$ stands for the volume element of the $S^{5}$.
The first equality in (\ref{eq:metricofmetallic}) just shows our notation that $\hat{g}_{\mu \nu}$ denotes the metric of the background geometry.
The metric of the $S^5$ is given by
\eq{
	d \Omega^2_5
&=
	d\theta^2
	+
	\sin^2 \theta
	d\psi^2
	+
	\cos^2 \theta
	d\Omega^2_3
,
\label{s5-metric}
}
where $\theta$ runs from 0 to $\pi/2$, and $\psi$ varies between $0$ and $2\pi$.  $d\Omega_3$ denotes the volume element of unit $S^3$.

The quark sector is mapped to D7-branes on the above geometry. In our study, we consider the case of single flavor, and we introduce only a single D7-brane. We employ the probe approximation since we are taking the large-$N_{c}$ limit.
The D7-brane wraps the $S^3$ part of the $S^5$ in such a way that the radius of the $S^3$ part depends on the radial coordinate $z$ in general. For the original proposal and the details on the D3-D7 system, see~\cite{Karch:2002sh}.

Let us exhibit the Dirac-Born-Infeld (DBI) action of the D7-brane for the purpose of defining our notations: 
\eq{
	S_{D7}
=
	-
	T_{D7}
	\int d^8 \xi
	\sqrt{-\det
	\roundLR{g_{ab}+2\pi\alpha' F_{ab}}
	}
,
\label{generalformofthedbiaction}
}
where $g_{ab}=\partial_a X^\mu \partial_b X^\nu \hat g_{\mu \nu}$ is the induced metric, $T_{D7}$ is the D7-brane tension, $\xi^{a}$ are the D7-brane's worldvolume coordinates,
$X^\mu$ represents the location of the D7-brane, and $F_{ab}=\partial_a A_b - \partial_b A_a$ is the field strength of the worldvolume U(1) gauge field ($a,\, b$ are the worldvolume indices). In our convention, $\alpha'^{-2}=4 \pi g_s N_c = 2g_{YM}^2 N_c = \lambda$, where $g_{s}$ is the string coupling and $g_{YM}$ is the gauge coupling constant.
The Wess-Zumino term will not contribute in our analysis, and the dilaton is trivial in our background geometry.
In this paper, we employ the static gauge where $\xi^a = \roundN{t, \vec{x}, z, \vec{\Omega}_3}$.

We assume the translational invariance along the $\vec{x}$ directions and the rotational invariance on the $S^3$.
Furthermore, we assume the configuration of the D7-brane is time independent.
We introduce the chemical potential for the quark charge, that corresponds to the boundary value of $A_t$ given at (\ref{eq:solofAtandh}), into this system~\cite{Nakamura:2006xk,Kobayashi:2006sb}. We apply the external electric field $E$ along the $x^1$ direction, which is encoded in the vector potential as $A_x(t,z)=-Et+h(z)$ within our gauge choice and ansatz \cite{Karch:2007pd}. We also employ an ansatz $\psi={\rm const.}$ $(\partial_{z}\psi = 0)$ which is consistent with the equation of motion of $\psi$.

Then the action per unit volume, which we redefine as $S_{D7}$, is now given as follows:
\eq{
	S_{D7}
&=
	-
	\mathcal{N}
	\int d t d z
	\cos^3 {\theta}
	{g}_{xx}
	\sqrt{
	\absLR{{g}_{tt}}
	{g}_{xx}
	{g}_{zz}
	-
	(2\pi\alpha')^2
	\roundLR{{g}_{xx} {{A_t}'}^2 + {g}_{zz} {\dot{A}_x}^2 - \absLR{{g}_{tt}} {A_x'}^2}
	}
,
\label{action-density}
}
where $x$ denotes $x^1$, and the dot and the prime stand for $\partial_t$ and $\partial_z$, respectively.
We have integrated over the $S^{3}$ directions in the derivation of (\ref{action-density}), and 
$
	\mathcal{N}
=
	2\pi^2
	T_{D7}
=
	\frac{\lambda}{(2\pi)^4}
	N_c
$
in our convention. Within our ansatz and gauge, the induced metric is
the same as the bulk metric except for the $g_{zz}$ component:
\eq{
	{g}_{tt}
&=
	-
	\frac{1}{z^2}
	\frac{(1 - z^4/z^4_H)^2}{1 + z^4/z^4_H}
,
\quad
	{g}_{xx}
=
	\frac{1}{z^2}
	\roundLR{1 + \frac{ z^4}{z^4_H}}
,
\quad
	{g}_{zz} 
=
	\frac{1}{z^2}
	+
	{\theta'}^2
.
}

\subsection{Gauge field}
Let us remind ourselves of the analysis given in \cite{Karch:2007pd}.
The equations of motion for the gauge field are
\eq{
	\frac{\partial \mathcal{L}}{\partial { A}_t'}
&=
	\cos^3 \theta
	{ g}_{xx}
	\frac{
	-
	\mathcal{N}
	(2\pi\alpha')^2
	{ g}_{xx}
	{ A}_t'
	}{
	\sqrt{
	\absLR{{ g}_{tt}}
	{ g}_{xx}
	{ g}_{zz}
	-
	(2\pi\alpha')^2
	\roundLR{
	{ g}_{xx}
	{ A_t'}^2
	+
	{ g}_{zz}
	\dot { A_x}^2
	-
	\absLR{{ g}_{tt}}
	{ A_x'}^2
	}
	}
	}
\equiv
	D
,
\\
	\frac{\partial \mathcal{L}}{\partial {A_x}'}
&=
	\cos^3 \theta
	{ g}_{xx}
	\frac{
	\mathcal{N}
	(2\pi\alpha')^2
	\absLR{{ g}_{tt}}
	h'
	}{
	\sqrt{
	\absLR{{ g}_{tt}}
	{ g}_{xx}
	{ g}_{zz}
	-
	(2\pi\alpha')^2
	\roundLR{
	{ g}_{xx}
	{ A_t'}^2
	+
	{ g}_{zz}
	\dot { A_x}^2
	-
	\absLR{{ g}_{tt}}
	{ A_x'}^2
	}
	}
	}
\equiv
	J
\label{eq:metallicEOMofA}
,
}
where $D$ and $J$ stand for the integral constants.
From (\ref{eq:metallicEOMofA}), we obtain
\eq{
	{ g}_{xx}
	{ A_t'}^2
&=
	\frac{1}{(2\pi\alpha')^2}
	\absLR{{ g}_{tt}}
	D^2
	\frac{
	{ g}_{zz}
	\roundLR{
	\absLR{{ g}_{tt}} 
	{ g}_{xx} 
	-
	(2\pi \alpha')^2 
	E^2
	}
	}{
	\mathcal{N}^2
	(2\pi \alpha')^2
	\absLR{{ g}_{tt}} 
	{ g}_{xx}^3
	\cos^6 { \theta}
	+
	\absLR{{ g}_{tt}} 
	D^2
	-
	{ g}_{xx}
	J^2
	}
,
\\
	\absLR{{ g}_{tt}}
	{h'}^2
&=
	\frac{1}{(2\pi\alpha')^2}
	{ g}_{xx}
	J^2
	\frac{
	{ g}_{zz}
	\roundLR{
	\absLR{{ g}_{tt}} 
	{ g}_{xx} 
	-
	(2\pi \alpha')^2 
	E^2
	}
	}{
	\mathcal{N}^2
	(2\pi \alpha')^2
	\absLR{{ g}_{tt}} 
	{ g}_{xx}^3
	\cos^6 { \theta}
	+
	\absLR{{ g}_{tt}} 
	D^2
	-
	{ g}_{xx}
	J^2
	}
.
\label{eq:solutionofAinmetallic}
}
At the vicinity of the boundary, the gauge fields can be expanded as 
\eq{
	{ A}_t(z)
&=
	\mu
	- 
	\frac{1}{2}
	\frac{D}{\mathcal{N}(2\pi \alpha')^2}
	z^2
	+
	O(z^4)
,
\\
	h(z)
&=
	b
	+
	\frac{1}{2}
	\frac{J}{\mathcal{N}(2\pi \alpha')^2}
	z^2
	+
	O(z^4)
.
\label{eq:solofAtandh}
}
The leading (non-normalizable) terms give the sources for the conjugate operators. 
${A}_t$ is conjugate to the charge density $J_t$; hence we interpret $\mu$ as the chemical potential. 
As is discussed in \cite{Kobayashi:2006sb} we require ${ A}_t(z_H) = 0$ which then fixes $D$ in terms of $\mu$.
For $h(z)$ we demand simply that the source term $b$ vanishes.
The subleading (normalizable) terms give the expectation values of the conjugate operators,
$
	\angleN{J^t}
=
	D
$,
$	\angleN{J^x}
=
	J
$
following the the Gubser-Klebanov-Polyakov-Witten relation.

The on-shell DBI action is now given by
\eq{
	S_{D7}
&=
	-
	\mathcal{N}
	\int dz dt
	\cos^6 { \theta}
	{ g}_{xx}^{5/2}
	\absLR{{ g}_{tt}}^{1/2}
	\sqrt{
	\frac{
	{ g}_{zz}
	\roundLR{  \absLR{{ g}_{tt}} { g}_{xx} - (2\pi \alpha') ^2 E^2}
	}{
	\absLR{{ g}_{tt} }
	{ g}_{xx}^{3} \cos^6 { \theta}
	+
	\frac{ \absLR{{ g}_{tt} } D^2 - { g}_{xx} J^2 }{ \mathcal{N}^2 (2\pi \alpha')^2 }
	}
	}
,
\label{eq:dbiactionwithsols}
}
which can be complex in general. However, we are studying the steady states, and we request the DBI action to be real. This requires that the term in the square root be positive semidefinite for all regions of $0 \leq z\leq z_H$:
\eq{
	\frac{
	{ g}_{zz}
	\roundLR{  \absLR{{ g}_{tt}} { g}_{xx} - (2\pi \alpha') ^2 E^2}
	}{
	\absLR{{ g}_{tt} }
	{ g}_{xx}^{3} \cos^6 { \theta}
	+
	\frac{ \absLR{{ g}_{tt} } D^2 - { g}_{xx} J^2 }{ \mathcal{N}^2 (2\pi \alpha')^2 }
	}
&\geq 
	0
,
\label{eq:termsinsquareroot}
}
which is achieved by setting both the numerator and the denominator to flip signs at the same point, say $z=z_*$, between $z=0$ and $z_H$~\cite{Karch:2007pd}.
Then the reality condition is reduced to 
\begin{eqnarray}
	\absLR{{ g}_{tt}} { g}_{xx} - (2\pi \alpha') ^2 E^2
	\Big|_{z=z_*}
&=
	0
,
\label{eq:definitionofzstar1}
\\
	\absLR{{ g}_{tt} }
	{ g}_{xx}^{3} \cos^6 { \theta}
	+
	\frac{ \absLR{{ g}_{tt} } D^2 - { g}_{xx} J^2 }{ \mathcal{N}^2 (2\pi \alpha')^2 }
	\Bigg|_{z=z_*}
&=
	0
.
\label{eq:definitionofzstar2}
\end{eqnarray}
From (\ref{eq:definitionofzstar1}), we have
\eq{
	z_*^2
&=
	\roundLR{ \sqrt{e^2 + 1} - e }
	z_H^2
,
\qquad
	e
\equiv
	\frac{ \absLR{E}}{\frac{\pi}{2} \sqrt{\lambda} T^2}
,
}
where $e$ is a dimensionless quantity.
Then (\ref{eq:definitionofzstar2}) gives
\eq{
	J^2
&=
	\roundLR{
	\frac{N_f^2 N_c^2 T^2}{16 \pi^2}
	\sqrt{e^2 + 1}
	\cos ^6 { \theta}(z_*)
	+
	\frac{d^2}{e^2+1}
	}
	E^2
,
\label{eq:RelaBetBandeanddandE}
}
where the dimensionless quantity $d$ has been defined as
\eq{
	d
&\equiv
	\frac{D}{\frac{\pi}{2} \sqrt{\lambda} T^2}
=
	\frac{ \angleLR{J^t} }{\frac{\pi}{2} \sqrt{\lambda} T^2}	
.
}
From (\ref{eq:RelaBetBandeanddandE}) we obtain the nonlinear conductivity $\sigma$ as~\cite{Karch:2007pd}
\eq{
	J
=
	\sigma
	E
,
\qquad
	\sigma
&\equiv
	\sqrt{
	\frac{N_f^2 N_c^2 T^2}{16 \pi^2}
	\sqrt{e^2 + 1}
	\cos ^6 { \theta}(z_*)
	+
	\frac{d^2}{e^2+1}
	}
.
}

\subsection{Scalar field}
The Euler-Lagrange equation for $\theta$ from (\ref{action-density}) is coupled to the gauge field.
Substituting (\ref{eq:solutionofAinmetallic}) into the equation of motion, it is given by
\eq{
&	\partial_z
	\squareLR{
	\frac{
	 { \theta} '
	}{
	(2\pi \alpha') 
	 }
\sqrt{	\frac{
   \left((2\pi \alpha')^2 E^2+g_{xx}
   { g}_{tt}\right)
	\roundLR{
	(2\pi \alpha')^2 \mathcal{N}^2 g_{xx}^3
   { g}_{tt} \cos ^6{ \theta}+J^2
   g_{xx}+D^2 { g}_{tt}
	}	
}{-{ g}_{tt} g_{xx} 
   g_{zz}   }
}   	}
\\
&
-
	3
	(2\pi \alpha')
	\mathcal{N}^2
	g_{xx}^{5/2} 
	\roundLR{ - { g}_{tt} { g}_{zz}}^{1/2}
	\sin { \theta} \cos ^5{ \theta}
   \sqrt{
	\frac{  
   (2\pi \alpha')^2
   E^2+g_{xx} { g}_{tt}
	}{
	(2\pi \alpha')^2 \mathcal{N}^2 g_{xx}^3
   { g}_{tt} \cos ^6{ \theta}+J^2
   g_{xx}+D^2 { g}_{tt}
	}
	}
=
	0
.
}

The asymptotic solution at the boundary is
\eq{
	\theta = \theta_0 z + \theta_2 z^3 + \cdots
,
}
where $\theta_0$ is related to the current quark mass $M_q$ of the fundamental representation, the mass of the charge carrier, as $M_q = \frac{1}{2}\sqrt{\lambda}T \theta_0$. $\theta_2$ gives the quark condensate as
$
	\angleLR{\bar{\psi}\psi}
=
	-
	\frac{1}{8}
	\sqrt{\lambda}
	N_c
	T^3
	\theta_2
$ \cite{Kobayashi:2006sb}.

\begin{figure}[h]
	\begin{minipage}[t]{0.48\linewidth}
		\includegraphics[width=\linewidth]{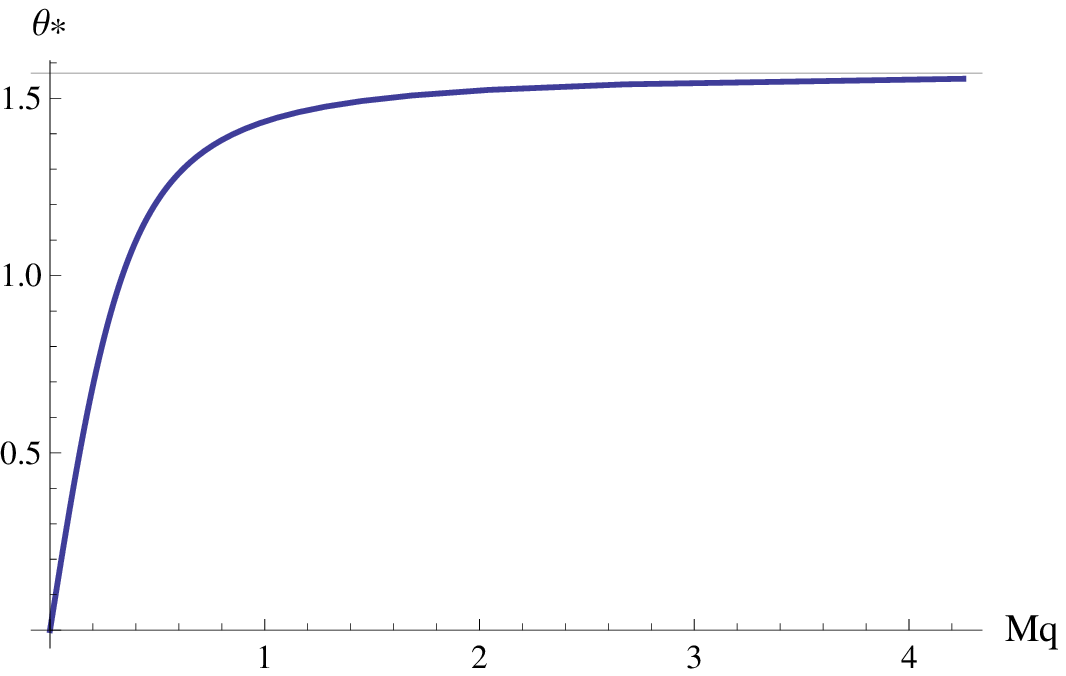}
	\end{minipage}
	\hspace{0.01\linewidth}
	\begin{minipage}[t]{0.48\linewidth}
		\includegraphics[width=\linewidth]{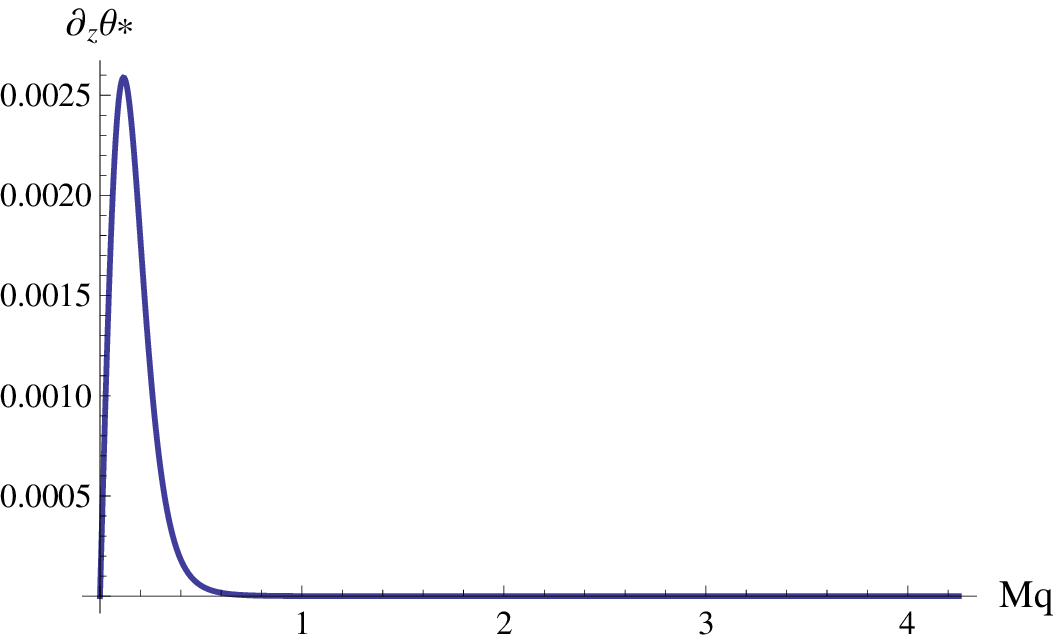}
	\end{minipage}
	\caption{$\theta(z_*)$ and $\partial_z \theta(z_*)$ as functions of $M_q$. 
 The curves are computed at $T=0.1$, $E=0.1$ and $D^2=0.1$. The straight line in the left figure indicates the asymptotic value $\theta(z_*)=\pi/2$. 
	}	
\label{fig:mqvsthetastar}
\end{figure}

Here, we proceed further than \cite{Karch:2007pd}, by investigating into the relationship between $\theta(z_*)$ and $M_q$ in detail for later use.
The reason why we are interested in the relationship between $M_q$ and $\theta'(z_*)$ but not $\theta'(z_H)$, is that $z=z_*$ turns out to be the location of ``effective horizon" in Section~\ref{diagonalization}.
The equation of motion for $\theta(z)$ at $z=z_{*}$ is given by
\eq{
\theta '
\squareLR{
(2\pi \alpha')^2 \mathcal{N}^2 g_{xx}^3
   g_{tt} \cos ^6\theta+J^2
   g_{xx}+D^2 g_{tt}
}'
+
3
(2\pi \alpha') ^2
\mathcal{N}^2
g_{xx}^{3} 
g_{tt}
g_{zz}
\sin \theta \cos ^5\theta
&=
0,
\label{NHeq}
}
which relates $\theta^{\prime}(z_*)$ to $\theta(z_*)$: the boundary condition at $z=z_{*}$ is given once we specify $\theta(z_*)$. Then we can solve the equation of motion numerically to find $M_q$ from the boundary value of $\theta$.
Fig.~\ref{fig:mqvsthetastar} demonstrates the behaviors of $\theta(z_*)$ (we may write $\theta_{*}$ as an abbreviation of $\theta(z_*)$) and $\partial_z \theta(z_*)$ as functions of $M_q$ 
at $T=0.1$, $E=0.1$ and $D^2=0.1$.\footnote{In the numerical computations, we set $2\pi \alpha^{\prime}={\mathcal N}=1$ for simplicity.}
We find that $\theta(z_*)$ is a monotonically increasing function of $M_q$ starting from zero at $M_q=0$ and approaching to $\pi/2$ when $M_q \rightarrow \infty$. We also find $ \theta'(z_*)=0$ at $M_q=0,\, \infty$. One finds that $ \theta'(z_*)=0$ at $|D|= \infty$ from (\ref{NHeq}) as well.

\section{Fluctuations and effective temperature}
\label{fluctuations}

The main purpose of the present work is to investigate the properties of the effective temperature of NESS. Of course, the notion of temperature in nonequilibrium systems is debatable. In our paper, we define the effective
temperature from the relationship between the small fluctuations of physical quantities and the corresponding dissipations~\cite{effectiveT,Gursoy:2010aa,Nakamura:2013yqa}. Therefore, analysis of small fluctuations is essential in defining the effective temperature in our study.

The fluctuations of physical quantities correspond to the fluctuations of normalizable modes in the gravity dual. Hence we are most interested in the equations of motion of fluctuations on the probe brane around the background configuration corresponding to NESS.

\subsection{Effective metric}
\label{effectiveM}

Let us consider the fluctuations of $X_\mu$ and $A_a$ (which we write $\tilde{X}^\mu$ and $\tilde{A}^a$, respectively) around the solutions obtained in Section \ref{setup} (which we write $\bar{X}^\mu$ and $\bar{A}^a$, respectively). 
The equations of motion of $\tilde{X}^\mu$ and $\tilde{A}^a$ are given by perturbing the equations of motion of $X^\mu$ and $A^a$ with the replacement $X^\mu \rightarrow \bar{X}^\mu + \tilde{X}^\mu$ and $A^a \rightarrow \bar{A}^a + \tilde{A}^a$. 

It is worthwhile to mention for arbitrary setups, and let
us begin with the DBI action of a D$p$-brane on an arbitrary background geometry whose metric is $\hat{g}_{\mu\nu}$:
\begin{eqnarray}
S=-T_{p}\int d^{p+1}\xi\: e^{-\Phi}\sqrt{-\det(g_{ab}+2\pi\alpha^{\prime}F_{ab})},
\end{eqnarray}
where $T_{p}$ is the tension of the D$p$-brane, $\xi^{a}$ are the worldvoulme coordinates, $\Phi$ is the dilaton field,
$g_{ab}=\partial_{a}X^{\mu}\partial_{b}X^{\nu}\hat{g}_{\mu\nu}$ is the induced metric, and $F_{ab}=\partial_{a}A_{b}-\partial_{b}A_{a}$ is the field strength of the worldvolume $U(1)$ gauge field.
The equations of motion of $X^\mu$ and $A^a$ are\footnote{See, for example, Appendix A of~\cite{Hashimoto:2014yza}.} 
\begin{eqnarray}
-\partial_{b}
\left(e^{-\Phi}\omega \sqrt{-G}\hat{g}_{\mu\nu}G^{ab}\partial_{a}X^{\mu}\right)
+
\frac{1}{2}
e^{-\Phi}
\omega\sqrt{-G}G^{ab}
\partial_{\nu}\hat{g}_{\alpha\beta}
\partial_{a}X^{\alpha}\partial_{b}X^{\beta}
&=&0,
\label{eomX}
\\
\partial_{a}\left(
e^{-\Phi}
\omega\sqrt{-G}G^{ab}F_{bc}g^{cd}
\right)&=&0,
\label{eomA}
\end{eqnarray}
where 
\begin{eqnarray}
G_{ab}=g_{ab}-(2\pi\alpha^{\prime})^{2}(Fg^{-1}F)_{ab}
\end{eqnarray}
is the open-string metric~\cite{Seiberg:1999vs,Gibbons:2000xe} and $\omega=(g/G)^{1/4}$. 
Note that $\Phi$ and $\hat{g}_{\mu\nu}$ contain $X^{\mu}$; $g_{ab}$ contains $X^{\mu}$ and $\partial_{a}X^{\mu}$; $\omega$ and $G_{ab}$ contain $X^{\mu}$, $\partial_{a}X^{\mu}$ and $F_{ab}$ in general. They may provide nontrivial interactions.

Now, we substitute $X^\mu = \bar{X}^\mu + \tilde{X}^\mu$ and $A^a = \bar{A}^a + \tilde{A}^a$ into (\ref{eomX}) and (\ref{eomA}), and consider the equations of motion for $\tilde{X}^\mu$ and $\tilde{A}^a$ to the {\em linear order} in fluctuations. The equations of motion can be divided into groups of (i) the terms with second derivative of fluctuations and (ii) the terms with first derivative or without derivative of fluctuations.
In Section \ref{diagonalization}, we find that $z=z_{*}$ plays the role
of a horizon of the geometry whose metric is Gab. At the horizon, the terms of (i) become dominant because of the redshift and the terms of (ii) are negligible. Therefore, if we are interested in the behavior of the fluctuations at the vicinity of $z=z_{*}$, we need only the terms of i)~\cite{Kinoshita}. The reason why we focus on the vicinity of $z=z_{*}$ shall be explained shortly.

Of course, the foregoing argument can be justified only when the fluctuations indeed obey the equations of motion on a curved spacetime given by the metric $G_{ab}$. We show it is indeed the case, at least for some special cases. 
The terms of (i) above in the static gauge can be written as follows:
\begin{eqnarray}
e^{-\Phi}\omega
\partial_{b}
\left( \sqrt{-G}G^{ab}\partial_{a}\tilde{X}^{\perp}\right)
+ 
({\rm terms\ which\ contain\ } \partial_{a}\bar{X}^{\perp})
&=&0,
\label{eomXtilde}
\\
e^{-\Phi}\omega g^{cd}
\partial_{a}\left(
\sqrt{-G}G^{ab}\tilde{F}_{bc}
\right)
+ 
({\rm terms\ which\ contain\ } \partial_{a}\bar{X}^{\perp})
&=&0,
\label{eomAtilde}
\end{eqnarray}
where $\tilde{F}_{ab}=\partial_{a}\tilde{A}_{b}-\partial_{b}\tilde{A}_{a}$.
The dilaton, the induced metric and the open-string metric contain only the 
background solutions here.
$X^{\perp}$ denotes $X^{\mu}$ in the directions perpendicular to the worldvolume directions, which are the physical degrees of freedom in the static gauge. Therefore, for the cases with $\partial_{a}\bar{X}^{\perp}=0$,\footnote{For more general situations, we postpone the analysis in future work~\cite{workinprogress}. } (\ref{eomXtilde}) and (\ref{eomAtilde}) reduce to 
\begin{eqnarray}
\partial_{b}
\left( \sqrt{-G}G^{ab}\partial_{a}\tilde{X}^{\perp}\right)
&=&0,
\label{NHeomXtilde}
\\
\partial_{a}\left(
\sqrt{-G}G^{ab}\tilde{F}_{bc}
\right)
&=&0,
\label{NHeomAtilde}
\end{eqnarray}
which are the Klein-Gordon equation and the Maxwell equation, respectively, on a geometry whose metric is $G_{ab}$.

The reason why we are interested in the equations of motion at the vicinity of $z=z_{*}$ is that the computations of correlation functions of the fluctuations are governed by them in the following sense.
Since $z=z_{*}$ turns out to be a horizon (which we call the effective horizon) of the geometry given by the metric $G_{ab}$,
the ingoing-wave boundary condition for fluctuations has to be imposed at $z=z_{*}$.
This means that the correlation functions are parametrized by the Hawking temperature associated with the effective horizon rather than that at the bulk horizon $z=z_{H}$. Since both the fluctuations and the dissipations are
evaluated through the correlation functions, the effective
temperature defined by (a generalization of) the fluctuation-dissipation relation at NESS is given by the Hawking temperature of the effective horizon, but not the temperature of the heat bath. Now, (\ref{NHeomXtilde}) and (\ref{NHeomAtilde}) show that the effective temperature can be read from $G_{ab}$~\cite{Gubser:2006nz,CasalderreySolana:2007qw,Gursoy:2010aa,Kim:2011qh,Sonner:2012if,Nakamura:2013yqa}.

\subsection{Diagonalization of effective metric}
\label{diagonalization}
In our setup of the D3-D7 model, $G_{ab}$ is given by
\begin{eqnarray}
G_{ab}=g_{ab}
+
(2\pi\alpha^{\prime})^{2}
\left(
  \begin{array}{ccccc}
   \frac{F_{tz}^{2}}{g_{zz}}+\frac{E^{2}}{g_{xx}}    &  \frac{F_{tz}F_{xz}}{g_{zz}}  & 0   & 0   & \frac{E F_{xz}}{g_{xx}}   \\
 \frac{F_{tz}F_{xz}}{g_{zz}}    &  \frac{F_{xz}^{2}}{g_{zz}}+\frac{E^{2}}{g_{tt}}  &  0  & 0   & \frac{E F_{tz}}{-g_{tt}}   \\
   0   &  0  & 0   & 0   &  0  \\
   0    & 0   &  0  & 0  &  0  \\
   \frac{E F_{xz}}{g_{xx}}     &  \frac{E F_{tz}}{-g_{tt}}   & 0   &  0  & \frac{F_{tz}^{2}}{g_{tt}} +\frac{F_{xz}^{2}}{g_{xx}}  \\
  \end{array}
\right),
\end{eqnarray}
which has off-diagonal components owing to the non-vanishing field strength of the worldvoulme gauge field. 
In order to diagonalize this effective metric, we consider the following transformation for $t,x$ and $z$:
\eq{
	\begin{pmatrix}
	dt \\
	dx\\
	dz \\
	\end{pmatrix}
\longrightarrow 
	\begin{pmatrix}
	d\tau\\
	d\eta \\
	d\rho \\
	\end{pmatrix}
=
	\left(
	\begin{array}{c}
	 dt+\frac{ G_{xt} G_{xz}-G_{xx}
	   G_{tz}}{\roundLR{G_{xt}}^2-G_{xx} G_{tt}}dz \\
	 dx + \frac{ G_{xt}}{G_{xx}}dt  + \frac{ G_{xz}}{G_{xx}} dz \\
	 dz
	\end{array}
	\right)
,
}
and then the diagonalized metric $\mathcal{G}_{ab}$ is
\eq{
	\mathcal{G}_{\tau \tau}
&=
	-
	\frac{
	(2\pi \alpha')^2 \mathcal{N}^2 g_{xx}^2  \cos ^6\bar{\theta}
   \left(
	g_{xx}
   \absLR{g_{tt}}
   -
	(2\pi \alpha')^2 E^2
	\right)
	}{
	(2\pi \alpha')^2 \mathcal{N}^2 g_{xx}^3  \cos^6\bar{\theta}+D^2
	}
,
\\
	\mathcal{G}_{\rho \rho}
&=
	\frac{
	(2\pi \alpha')^2 \mathcal{N}^2 g_{xx}^3 \absLR{g_{tt}} g_{zz}  \cos
   ^6\bar{\theta}
	}{
	(2\pi \alpha')^2 \mathcal{N}^2 g_{xx}^3 \absLR{g_{tt}} \cos ^6\bar{\theta}
	+
	D^2 \absLR{g_{tt}}
	-
	J^2 g_{xx}
	}
,
\\
	\mathcal{G}_{\eta\eta}
&=
	\frac{
   \left((2\pi \alpha')^2 \mathcal{N}^2 g_{xx}^3  \cos ^6\bar{\theta}
   +D^2\right)
	\left(g_{xx} \absLR{g_{tt}} - (2\pi \alpha')^2 E^2\right)
	}{
	(2\pi \alpha')^2 \mathcal{N}^2 g_{xx}^3 \absLR{g_{tt}}
    \cos ^6\bar{\theta}
	+
	D^2
   \absLR{g_{tt}}
	-
	J^2 g_{xx}
	}
,
\\
	\mathcal{G}_{22}
&=
	\mathcal{G}_{33}
=
	g_{xx}
.
}
Note that the numerator of ${\mathcal G}_{\tau\tau}$ and the denominator of ${\mathcal G}_{\rho\rho}$ contain (\ref{eq:definitionofzstar1}) and (\ref{eq:definitionofzstar2}), respectively, which go to zero at $z=z_{*}$.
One can check that ${\mathcal G}_{\eta\eta}$ has nonzero and finite value at $z=z_{*}$.
Hence, near $z_*$, the effective metric 
behaves as follows:
\eq{
	\mathcal{G}_{\tau \tau}
&\sim
	-a(z-z_*)
,
\qquad
	\mathcal{G}_{\rho \rho}
\sim
	b/(z-z_*)
,
}
where
\eq{
	a
&=
	\frac{
	(2\pi \alpha')^2 \mathcal{N}^2 g_{xx}^2  \cos ^6\bar{\theta}
	\left( 
	g_{xx}
    \absLR{ g_{tt} } 
	\right)' 
	}{
	(2\pi \alpha')^2 \mathcal{N}^2 g_{xx}^3  \cos^6\bar{\theta}+D^2
	}
	\Bigg|_{z=z_*}
,
\qquad
	b
=
	\frac{\roundLR{2\pi \alpha' }^2 \mathcal{N}^2 g_{xx}^3 \absLR{ g_{tt} } g_{zz} \cos ^6\bar{\theta}}{	\roundLR{
	\frac{
	\roundLR{2\pi \alpha' }^2 \mathcal{N}^2 g_{xx}^3 \absLR{ g_{tt} }
	   \cos ^6\bar{\theta}+ D^2   \absLR{ g_{tt} }
	}{g_{xx}}
	}'
	g_{xx}
	}
	\Bigg|_{z=z_*}
.
\label{eq:aandb}
}
This means that $z=z_{*}$ plays the role of the horizon (effective horizon) for the small fluctuations of the normalizable modes on the probe D-brane when (\ref{NHeomXtilde}) and (\ref{NHeomAtilde}) hold.
Then the Hawking temperature associated with the effective horizon, which we call effective temperature $T_*$, can be read from the ratio of $a$ to $b$ as follows:
\eq{
	T_*
&=
	\frac{1}{4\pi}
	\sqrt{
	\frac{a}{b}
	}
.
\label{eq:Efftemperature}
}

\section{Results}
\label{results}

In this section, we present the results of our analysis on the effective temperature. First, we show two limiting cases where the results are obtained analytically, and then we present numerical results for more general cases.

In our setup, $X^{\perp}$ corresponds to $\theta$ and $\psi$. We have employed the ansatz (which is consistent with the equation of motion) $\partial_{a}\bar{\psi}=0$, and (\ref{NHeomXtilde}) always holds for $\tilde{\psi}$. We can also show that $\tilde{\psi}$ decouples from the other modes within the consideration of Section \ref{effectiveM}. 
Therefore, the notion of the effective temperature for $\tilde{\psi}$ is valid for all the cases presented in this section. For $\theta$, we have found in Section \ref{setup} that $\partial_{a} \bar{\theta}$ at $z=z_{*}$ vanishes when $M_{q}=0$, $M_{q}=\infty$ and when $|D|=\infty$. 
At these limits,  (\ref{NHeomXtilde}) for $\tilde{\theta}$ holds, and $\tilde{F}_{ab}$ obeys to (\ref{NHeomAtilde}).
Therefore, the results for these three limits given in Section \ref{resultslimit} are valid for all the physical fluctuations
$\tilde{\psi}$, $\tilde{\theta}$ and $\tilde{F}_{ab}$.

\subsection{Infinite-mass limit and high-density limit}
\label{resultslimit}

$T_*$ depends on $\bar\theta(z_*)$ as we can see in (\ref{eq:aandb}) and (\ref{eq:Efftemperature}). The relationship between $\bar\theta(z_*)$ and physical parameters, such as $M_{q}$, is obtained from the nonlinear equation of motion which is solvable only numerically in general.
However, we find that the large mass limit $M_q \rightarrow \infty$ corresponds to the limit of $\bar \theta(z_*) \rightarrow \pi/2$ in Fig.~\ref{fig:mqvsthetastar}, and $T_*$ at this limit can be computed analytically by using this property.
We find that the effective temperature behaves as
\eq{
	T_*
&=
	\frac{1}{4\pi}
	\sqrt{
	\frac{64}{ z_H^2 \sqrt{  4 + (2\pi \alpha')^2 E^2
   z_H^4}}
   +
   \mathcal{O} \roundLR{ \theta_* - \pi/2 }
   }
\longrightarrow  
	\frac{T}{\sqrt[4]{1 +e^2}}
,
\label{eq:Tstarinlargemasslimit}
}
at the large-mass limit.
Note that the effective temperature is {\em lower} than the heat-bath temperature at finite $E$.

Let us compare the effective temperature (\ref{eq:Tstarinlargemasslimit})
to that of the Langevin system given in \cite{Gubser:2006nz}.
In \cite{Gubser:2006nz}, the effective temperature is given as
$
	T_*
=
	\sqrt[4]{1-v^2}\,
	T
,
$
where $v$ stands for the velocity of the test quark.
In our system, the average velocity of the charge carriers is given by the following relationship:
\eq{
	v^2
&=
	\roundLR{
	\frac{J}{D}
	}^2
=
	\frac{e^2 }{1+ e^2 }  
.
\label{eq:relabetvandEandT}
}
The evaluation of the average velocity is justified since the contribution of the pair creation is absent at the large-mass limit. In the presence of the pair creation, the positively charged particles and the negatively charged particles are moving in opposite directions, and $J$ does not necessarily reflect the average velocity of the charge carriers.
Then we obtain from (\ref{eq:Tstarinlargemasslimit}) and (\ref{eq:relabetvandEandT})
\eq{
	T_*
&
=
	\sqrt[4]{1- v^2}
	~T
,
\label{eq:TsGubserCase}
}
which completely agrees with the result of the Langevin system of an infinitely heavy single test particle~\cite{Gubser:2006nz}.

One finds that we can also take the high-density limit, $D^2 \rightarrow \infty$, in (\ref{eq:aandb}) and (\ref{eq:Efftemperature})
 analytically. We obtain
\eq{
	T_*
&=
	\frac{1}{4\pi}
	\sqrt{
	\frac{64}{z_H^2 \sqrt{ 4 + (2\pi \alpha')^2 E^2
   z_H^4}}
   +
   \mathcal{O} \roundLR{ \frac{1}{D^2} }
   }
\longrightarrow 
	\frac{T}{\sqrt[4]{1+e^2}}
,
\label{eq:efftempininfinitedensitylimit}
}
which coincides with (\ref{eq:Tstarinlargemasslimit}). 
At the large-density limit, the contribution of the doped carriers dominates over that of the pair creation, and we can again justify the estimation (\ref{eq:relabetvandEandT}). Although we obtain (\ref{eq:efftempininfinitedensitylimit}) for arbitrary mass, it coincides with (\ref{eq:TsGubserCase}) at the large-density limit.

\subsection{Numerical results}

Numerical computation is necessary for the cases of arbitrary density and arbitrary mass. We show the results from the numerical analysis. We set $2\pi \alpha^{\prime}={\mathcal N}=1$ for simplicity in the numerical computations.

\begin{figure}[h]
	\begin{minipage}[t]{0.49\linewidth}
		\includegraphics[width=\linewidth]{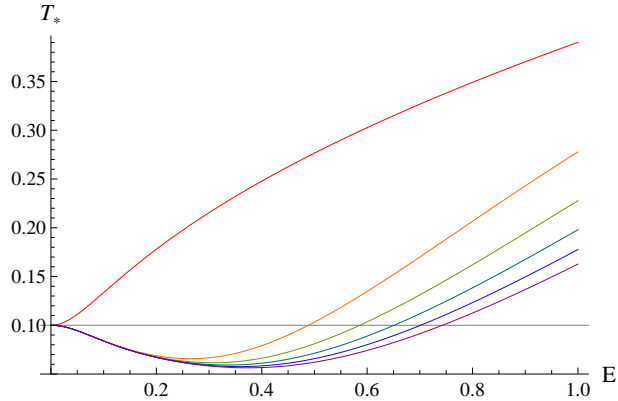}
		\caption{$T_*$ vs $E$ for massless case at $T=0.1$. The straight line indicates the temperature of the heat bath. The other curves correspond to different $D$'s from the upper curve ($D^2=0$) to the lower curve ($D^2=5$) in increments of $1$.}
		\label{fig:massless}
	\end{minipage}
\end{figure}

Fig.~\ref{fig:massless} shows the effective temperature at the massless limit but for various densities.
Here we set the heat-bath temperature to be $T=0.1$ which is indicated by the straight line for reference.  
We have checked that the result at $D=0$ agrees with that in \cite{Nakamura:2013yqa} where $T_* > T$ at finite $E$.
However, we find that a region of $T_* < T$ appears for $D\neq 0$ when $E$ is small but nonzero.
The condition for $T_* < T$ shall be found to be $D^2/T^6 > \lambda N_{c}^{2} /32$ at (\ref{eq:conditionforTstislowerthanT}).

\begin{figure}[h]
	\begin{minipage}[t]{0.48\linewidth}
		\includegraphics[width=\linewidth]{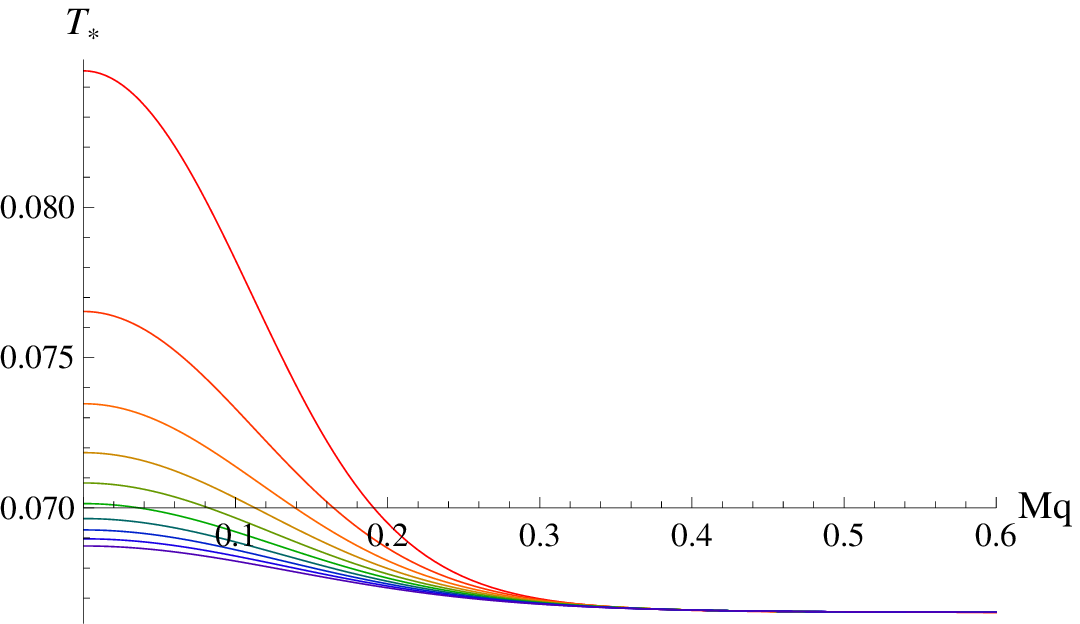}
		\caption{$T_*$ vs $M_{q}$ at $T=0.1$ and $E=0.2$. The curves correspond to different $D$'s from the upper curve ($D^2=0.1$) to the lower curve ($D^2=1$) in increments of $0.1$.}
		\label{fig:Teffvsthetasforfinitemass}
	\end{minipage}
	\hspace{0.01\linewidth}
	\begin{minipage}[t]{0.48\linewidth}
		\includegraphics[width=\linewidth]{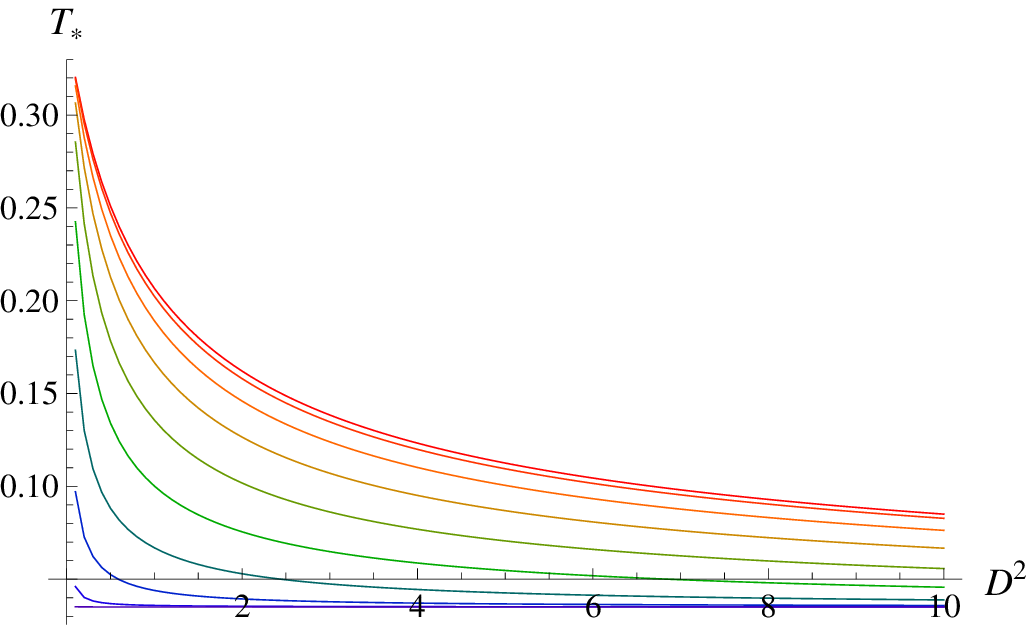}
		\caption{$T_*$ vs $D^2$ for various mass at $T=0.1$ and $E=0.8$. The curves correspond to different $\bar{\theta}(z_*)$'s from the upper curve ($\bar{\theta}(z_*)=0$) to the lower curve ($\bar{\theta}(z_*)=0.9 \times \pi/2$) in increments of $0.1 \times \pi/2$.}
		\label{fig:TeffvsD2forfinitemass}
	\end{minipage}
\end{figure}

The results for finite mass are given in Fig.~\ref{fig:Teffvsthetasforfinitemass}. We present the relationship between $T_*$ and $M_{q}$
In Fig.~\ref{fig:Teffvsthetasforfinitemass}, we have set $T=0.1$ and $E=0.2$. The curves correspond to different $D$'s from the upper line ($D^2=0.1$) to the lower one ($D^2=1$) in increments of $0.1$.
One can check the consistency, that $T_*$ at $M_q$
is the same as that at $E=0.2$ in Fig.~\ref{fig:massless}, and the curves degenerates into the value given by (\ref{eq:Tstarinlargemasslimit}) in the large $M_q$ region.

\begin{figure}[h]
	\begin{minipage}[t]{0.48\linewidth}
		\includegraphics[width=\linewidth]{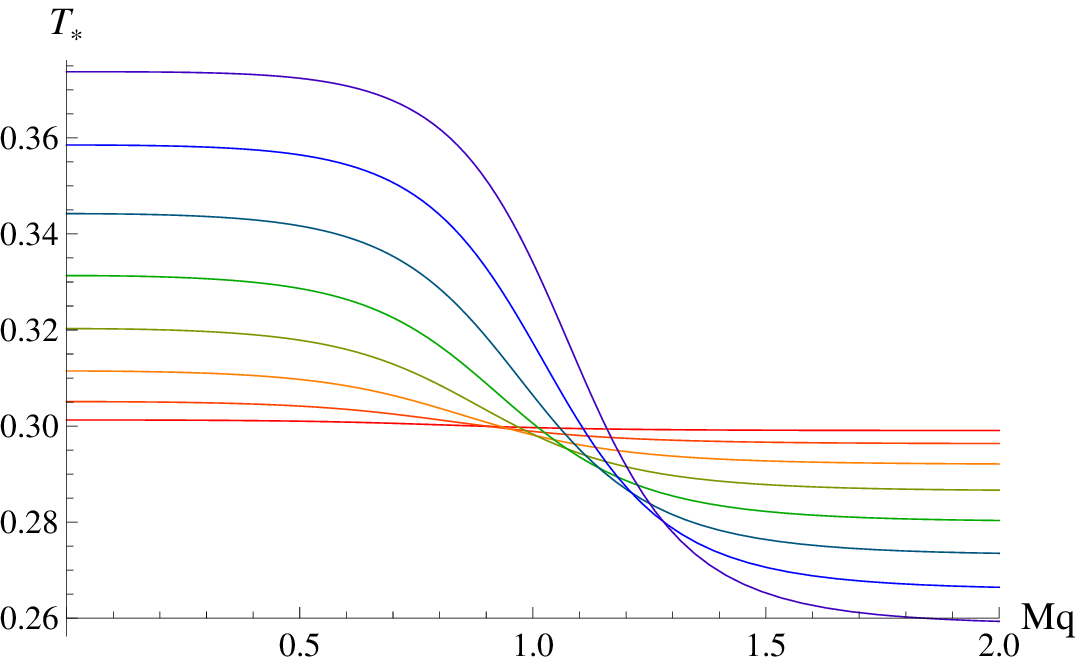}
		\caption{$T_*$ vs $M_{q}$ at $T=0.3$ and $D^2=0.2$. The curves correspond to different $E$'s.  It varies from $E=0.1$ to $E=0.8$ in increments of $0.1$ when we follow the intercept on the $T_*$ axis from up to down.
	}
		\label{fig:Teffvsthetasforfinitemass2}
	\end{minipage}
	\hspace{0.01\linewidth}
	\begin{minipage}[t]{0.48\linewidth}
		\includegraphics[width=\linewidth]{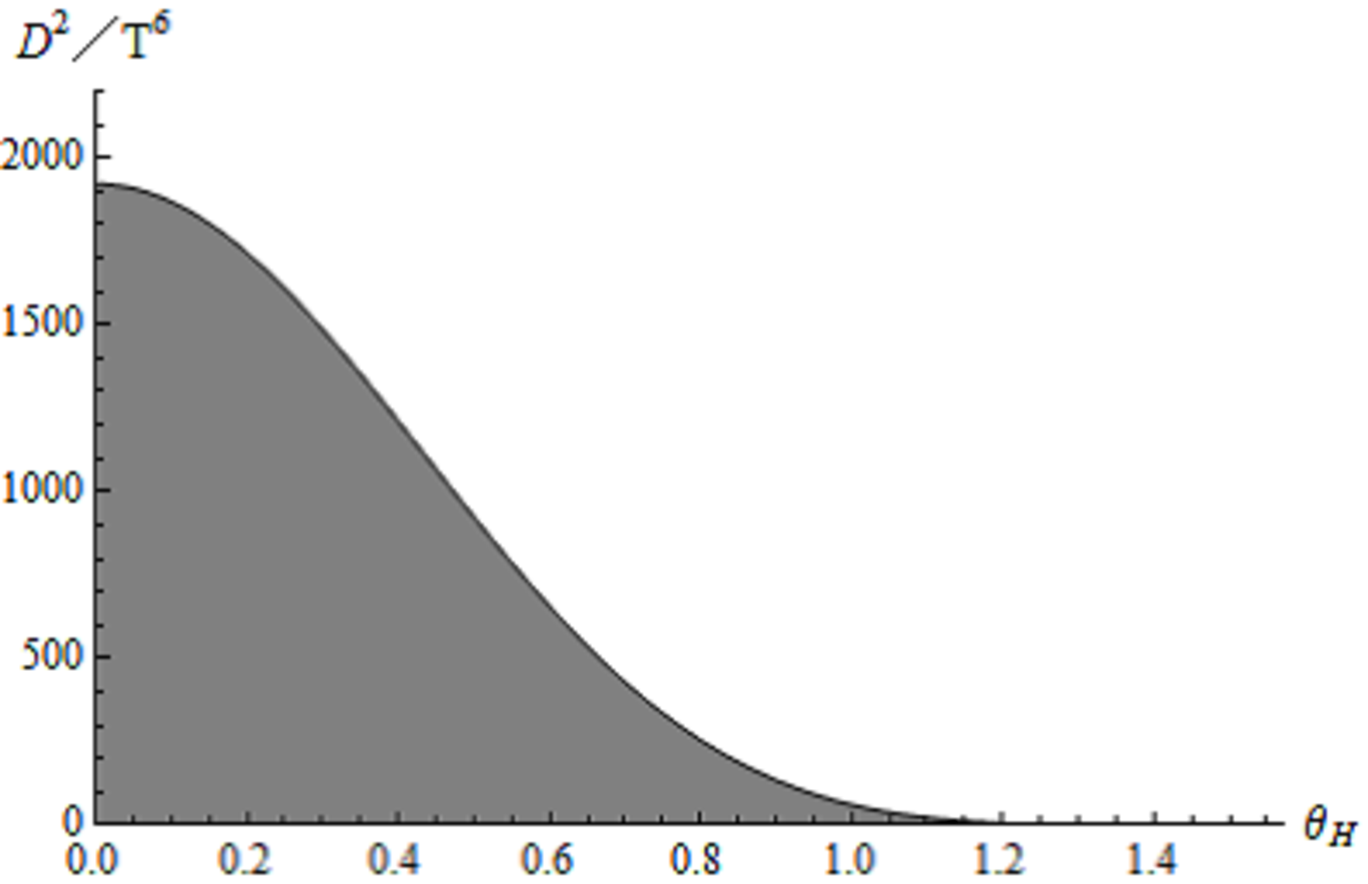}
		\caption{
A diagram for behavior of the effective temperature. The boundary line is $D^{2}/T^{6}=I$. The region under the line (filled by gray color) corresponds to the region for $D^{2}/T^{6}<I$ and hence $T_* > T$, whereas the region above the line (the white region) is for $D^{2}/T^{6}>I$ where $T_* < T$.}
		\label{fig:ConditionForTsLowerThanT}
	\end{minipage}
\end{figure}

Fig.~\ref{fig:TeffvsD2forfinitemass} shows $T_*$ vs $D^2$ at $T=0.1$ and $E=0.8$. The curves correspond to different $\bar{\theta}(z_*)$'s from the upper curve ($\bar{\theta}(z_*)=0$) to the lower one ($\bar{\theta}(z_*)=0.9 \times \pi/2$) in increments of $0.1 \times \pi/2$.
We present the relationship between $T_*$ and $\bar{\theta}(z_*)$ rather than that for $T_*$ and $M_{q}$, but we can read the dependence of $T_*$ on $M_{q}$ qualitatively since $M_{q}$ is a monotonically increasing function of $\bar{\theta}(z_*)$ as is demonstrated in Fig.~\ref{fig:mqvsthetastar}\footnote{We use $\bar{\theta}(z_*)$ rather than $M_q$ in Fig.~\ref{fig:TeffvsD2forfinitemass} and Fig.~\ref{fig:ConditionForTsLowerThanT} since $\bar{\theta}(z_*)=\pi/2$ corresponds to $M_q=\infty$, and $\bar{\theta}(z_*)$ is a useful parameter at the large $M_q$ region.}.
In this figure, the curves reach the same value given in (\ref{eq:efftempininfinitedensitylimit}) at high densities.
Independence of $T_{*}$ on $M_{q}$ at high densities can also be seen in
Fig.~\ref{fig:Teffvsthetasforfinitemass} where the dependence of $T_{*}$ on $\bar \theta(z_*)$ becomes weaker as $D$ goes large.

Fig.~\ref{fig:Teffvsthetasforfinitemass2} shows $T_*$ vs 
$M_q$ at $T=0.3$ and $D^2=0.2$. 
The curves correspond to different $E$'s.
It varies from $E=0.1$ to $E=0.8$ in increments of $0.1$ when we follow the intercept on the $T_*$ axis from up to down. The figure shows that $T_*$ of the system of light carriers
increases along $E$
\footnote{
Precisely speaking, this statement is correct when $|D|$ satisfies (\ref{eq:conditionforTstislowerthanT}).
},
whereas $T_*$ for the system of heavy carriers decreases along $E$.

This implies that the pair creation, which is dominant at small $M_q$, has an effect of raising the effective temperature, 
whereas the drag effect, which is dominant at large $M_q$, lowers the effective temperature.

\subsection{Small $E$ analysis}

So far we have found that the effective temperature becomes lower when the density and the mass of the carriers are large. In order to highlight this property, let us examine the effective temperature $T_*$ in the small $E$ region but for arbitrary density and mass.
Expanding $T_*$ with respect to $E$ to the order of $E^{2}$, we find that $T_* < T$ is realized when the following condition is satisfied:
\eq{
	\frac{D^2}{T^6}
&>
	I(\bar\theta_H)
,
\\
	I(\bar\theta_H)
&\equiv
\frac{\lambda N_{c}^{2}}{2^{9}}
	\cos ^{11/2}\bar\theta_H
	\squareB{
	4 
	\sqrt{
	\cos \bar\theta_H
	}
	+
	3 
	\sqrt{2} 
	\sqrt{
	4
	+
	7
	\cos \bar\theta_H
	-
	4
	\cos ( 2\bar\theta_H)
	+
	\cos ( 3\bar\theta_H )
	}
	}
,
\label{eq:conditionforTstislowerthanT}
}
where $\bar\theta(z_H)$ is abbreviated as $\bar\theta_H$.

Fig.~\ref{fig:ConditionForTsLowerThanT} shows the behavior of the effective temperature at small but nonzero $E$ for various densities and masses.
The region under the line (filled by gray color) corresponds to $D^2/T^6 < I(\bar \theta_H)$ and hence $T_* > T$, whereas the region above the line (the white region) is for $D^2/T^6 > I(\bar \theta_H)$ where $T_* < T$.

\subsection{General models}
\label{generalmodel}
We have mainly focused on the D3-D7 model so far, but the analysis can be straightforwardly generalized into other models. Suppose that our background geometry is the near-horizon geometry of the D$p$-brane solution~\cite{Itzhaki:1998dd} and our probe brane is a D$(q+n+1)$-brane. 
We assume that the probe brane wraps the $S^{n}$ part of the $S^{8-p}$ of the background geometry and extents in $(q+1)$-dimensional spacetime along the boundary directions as well as the radial direction. We apply the external electric field $E$ acting on the global charge along the $x$ direction.
This is almost the same setup as that is considered in \cite{Nakamura:2013yqa}, but we introduce the mass of the fundamental representations and the density of the global charge carried by the fundamental representations which have not been considered in  \cite{Nakamura:2013yqa}.

Since the analysis is parallel to what we have done in the D3-D7 model, we briefly present the main results. The effective temperature of the conductor system is given by
\eq{
T_*
&=
\frac{1}{4\pi}
\sqrt{
\frac{a}{b}
}
,
\\
a
&=
\frac{
w  g_{xx}^{q-1}
\cos^{2 n}\theta
   \left(
\absLR{g_{xx}} g_{tt}
\right)'
}{
w
   g_{xx}^q \cos ^{2 n}\theta+D^2
}
\Bigg|_{z_*}
,
\qquad
b
=
\frac{
w \absLR{g_{tt}} g_{zz} 
   g_{xx}^q \cos ^{2 n}\theta
}{
\roundLR{
\frac{
w
   \absLR{g_{tt}} g_{xx}^q \cos ^{2 n}\theta
   +D^2 \absLR{g_{tt}}}{g_{xx}}
   }'
   g_{xx}
}
\Bigg|_{z_*}
,
}
where $w$ is a model-dependent factor that includes the contributions of the dilaton, the tension of the probe brane, and the volume of the compact directions. $g_{ab}$ is the induced metric in the given setup.
Note that $T_{*}$ depends on $q$ and $n$ in general.
However, we find that $T_{*}$ becomes independent of $q$ and $n$ if we take the limit of $D \rightarrow \infty$ or $M_q \rightarrow \infty$ ($\theta_*=\pi/2$).
Furthermore, the effective temperatures coincide with each other at both limits, as is the case in the D3-D7 system. The effective temperature at these limits is
\eq{
	T_*
&=
	\frac{1}{4\pi}
	\sqrt{
	\frac{
	(\absLR{g_{tt}} g_{xx})'
	}{
	\absLR{g_{tt}} g_{zz}
	}
	\roundLR{
	\frac{\absLR{g_{tt}}}{g_{xx}}
   }'
	}
=
	\frac{T
	}{
	\left(
	1+e^2
	\right)^{\frac{1}{7-p}}
	}
=
	\left(
	1-v^2
	\right)^{\frac{1}{7-p}}
	T
,
}
where $e =	(2\pi \alpha') E
   \roundLR{
	\frac{7-p}{  4\pi T}
	}^{\frac{7-p}{5-p}}
$,
and the average velocity of the charge carrier is given as (\ref{eq:relabetvandEandT}).
The result, which is a generalization of (\ref{eq:Tstarinlargemasslimit}) and (\ref{eq:efftempininfinitedensitylimit}), agrees with the Langevin systems (dragged string cases) shown in \cite{Nakamura:2013yqa}.

\section{Discussions}
\label{discussions}
We have analyzed the properties of the effective temperature of NESS in holographic models. Our systems are many-body systems of charge carriers driven by the electric field. We find that at the large-density limit and at the large-mass limit of the charge carriers, the effective temperature agrees with that for the corresponding Langevin system.
Let us find possible interpretations of our result.

In the conductor systems, the charge carriers have two origins: those who have doped and those who have pair created. The pair creation is suppressed at the large-mass limit, and the effect of the doped carriers dominates at the large-density limit. This means that the effective temperature of the conductor systems and that of the Langevin systems agree when the role of the doped charge is dominant. This is natural in the sense that the systems of the doped carriers are the many-body systems of the single dragged particle in the same medium. 
However, our analysis shows more: the effective temperature of the {\em doped charges} is not affected by the interaction among them at the large-mass limit, since it is independent of the density at this limit. 
We also found that the effective temperature of the {\em doped charges} is not affected by the mass at the high-density limit, either. The reason why we have emphasized {\em doped charges} is that the effect of the pair creation becomes unimportant at these limits.
For the mutual consistency of these limits, the effective temperatures at these two limits have to agree with each other. We found it is indeed the case. 
Note that these properties are owing to neither the
supersymmetry of the microscopic theory nor the conformal
invariance, since we have observed the same
properties in general models that do not necessarily have
supersymmetry or conformal invariance\footnote{Note that the supersymmetry is broken in our setup because of the temperature and the density even if the original microscopic theory is supersymmetric.}.

We have also found that the effective temperature can be lower than the temperature of the heat bath even for the systems which show the higher effective temperature in the neutral case. For example, we found that the effective temperature of the D3-D7 system at finite densities can be lower than the heat-bath temperature in the region of small electric field. These observations lead us to the conclusion that the pair creation of charge carriers has an effect of raising the effective temperature, whereas dragging of the doped carriers lowers the effective temperature, in a wide range of holographic models of NESS we have studied.

It has been found that the systems we have studied show nonlinear conductivity, and some of them even show interesting characteristics such as negative differential conductivity~\cite{Nakamura:2010zd} and nonequilibrium phase transitions~\cite{Nakamura:2012ae}. It is interesting to see how the properties of the effective temperature contribute to these phenomena. We leave this for future study.

\section*{Acknowledgments}
We thank S. Kinoshita for fruitful discussions and comments. The work of S. N. was supported in part by the Grant-in-Aid for Scientific Research on Innovative Areas “Fluctuation \& Structure” (No. 26103517) and
Grant-in-Aid for Challenging Exploratory Research
(No. 23654132) from the Ministry of Education,
Culture, Sports, Science, and Technology of Japan; the
Chuo University Personal Research Grant; and the visiting
professorship of the Institute for Solid State Physics,
University of Tokyo.


\end{document}